# Dynamic anti-plane sliding of dissimilar anisotropic linear elastic solids

K. Ranjith

Associate Professor

Mechanical Engineering Department

Amrita University, Coimbatore 641105, Tamil Nadu, India

E-mail: ranjith@post.harvard.edu

Phone: +91-422-2656422 Ext. 523, Fax: 91-422-2652174

**Abstract**:

The stability of steady, dynamic, anti-plane slipping at a planar interface between two dissimilar anisotropic linear elastic solids with is studied. The solids are assumed to possess a plane of symmetry normal to the slip direction, so that in-plane displacements and normal stress changes on the slip plane do not occur. Friction at the interface is assumed to follow a rate and state dependent law with velocity-weakening behavior in the steady state. The stability to spatial perturbations of the form $\exp(ikx_1)$, where $k$ is the wavenumber and $x_1$ is the coordinate along the interface is studied. The critical wavenumber magnitude, $|k|_{cr}$, above which there is stability and the corresponding phase velocity, $c$, of the neutrally stable mode are obtained from the stability analysis. Numerical plots showing the dependence of $|k|_{cr}$ and $c$ on the unperturbed sliding velocity, $V_o$, are provided for various bi-material combinations of practical interest.

**Keywords:** stability; bifurcation; anisotropic; elasticity; anti-plane; interface; friction; slip; earthquakes; biological; technological

# 1. Introduction

The problem of stability of frictional sliding of solids in elastic surroundings is introduced below with several simple examples. The simplest is one wherein a rigid, massless block under a constant compressive normal stress $\sigma_o$ is pulled along a frictional surface by a spring of stiffness $K$ attached to it. A constant velocity $V_o$ is applied to one end of the spring and the sliding velocity of the block is $V$. At steady state, $V = V_o$. Consider a slip perturbation from the steady state. The elastic relation for the corresponding shear stress perturbation is then

$$\dot{\tau} = -K(V - V_o). \qquad (1)$$

We first study whether steady sliding is stable, i.e. $V \to V_o$, when friction is assumed to follow a purely rate dependent law, $\tau = \tau(V)$. Linearizing eqn. (1) about the steady state gives

$$\tau'(V_o) \frac{d(V - V_o)}{dt} + K(V - V_o) = 0 \qquad (2)$$

and it follows that steady sliding is stable if the surface is velocity strengthening, $\tau'(V_o) > 0$, and unstable if the surface is velocity weakening, $\tau'(V_o) < 0$. This however contradicts experimental evidence that sliding on a velocity weakening surface can be stable if the block is pulled with a sufficiently stiff spring. The conclusions regarding stability are unchanged if a non-zero mass is assumed for the block. The governing equation in that case becomes

$$m \frac{d^2 V}{dt^2} + \frac{d\tau(V)}{dt} + K(V - V_o) = 0 \qquad (3)$$

where $m$ is the mass of the block. Linearizing about the steady state with velocity $V = V_o$, eqn. (3) may be written as

$$m\frac{d^2(V-V_o)}{dt^2} + \tau'(V_o)\frac{d(V-V_o)}{dt} + K(V-V_o) = 0 \quad (4)$$

Clearly, the system is stable if $\tau'(V_o) > 0$ and unstable if $\tau'(V_o) < 0$.

The work of Dieterich (1979) and Ruina (1983) has, however, made a case for a more elaborate framework for the velocity dependence of friction. They show that at a constant normal stress, $\sigma_o$, friction depends not just on sliding velocity, but also on a state variable which represents fading memory of the history of sliding velocity. Thus

$$\tau = \tau(V,\theta), \quad (5)$$

where $\theta$ is a state variable. A common way to write this relation is

$$\tau = \tau_o + a\sigma_o \ln(V/V_o) + b\sigma_o \ln(V_o\theta/L) \quad (6)$$

where $\tau_o$ is the frictional stress at the steady sliding velocity $V_o$ and $a, b$ and $L$ are positive constants. Several forms are in use for the evolution of the state variable with the most common being the Dieterich-Ruina ageing law,

$$d\theta/dt = 1 - V\theta/L, \quad (7)$$

and the Ruina-Dieterich slip law,

$$d\theta/dt = -(V\theta/L)\ln(V\theta/L). \quad (8)$$

For steady sliding at a velocity $V$, both evolution laws give the steady state shear resistance as

$$\tau_{ss}(V) = \tau_o - \sigma_o(b-a)\ln(V/V_o). \quad (9)$$

Clearly, a surface is velocity weakening or velocity strengthening as is $(b-a)$ greater or less than zero. Linearizing eqn. (6) in the slip velocity about a steady sliding state at a velocity $V_o$ (keeping normal stress constant at $\sigma_o$), and eliminating the explicit dependence on the state variable, one gets (Rice, 1983)

$$\frac{d\tau}{dt} = \frac{a\sigma_o}{V_o}\frac{dV}{dt} - \frac{V_o}{L}[\tau - \tau_o + \frac{(b-a)\sigma_o}{V_o}(V - V_o)]. \quad (10)$$

Following Ruina (1983), consider quasi-static slip of the block, so that $\dot{\tau} = -K(V - V_o)$. Substituting that elastic relation in eqn. (10), we get

$$\frac{a\sigma_o}{V_o}\frac{d^2(V - V_o)}{dt^2} + \left[K - \left(\frac{b-a}{L}\right)\sigma_o\right]\frac{d(V - V_o)}{dt} + K\frac{V_o}{L}(V - V_o) = 0 \quad (11)$$

Thus quasi-static slip of a spring-block system with a rate and state dependent friction law is stable, i.e., $V \to V_o$, even for velocity weakening surfaces, $(b - a) > 0$, as long as the spring is stiffer than a critical value given by

$$K_{cr} = \frac{\sigma_o(b-a)}{L}. \quad (12)$$

Sliding is always stable with a velocity strengthening friction law. When $K = K_{cr}$, the oscillation frequency is clearly

$$\omega = \sqrt{K_{cr}\frac{V_o}{L}\frac{V_o}{a\sigma_o}} = \sqrt{\frac{b-a}{a}}\frac{V_o}{L} \quad (13)$$

and

$$V - V_o \; ; \; e^{\pm i\omega t}. \quad (14)$$

The corresponding result for $K_{cr}$ with the mass of the block taken into account is (Rice and Ruina, 1983)

$$K_{cr} = \frac{\sigma_o(b-a)}{L}\left[1 + \frac{mV_o^2}{a\sigma_o L}\right] \quad (15)$$

The oscillation frequency, $\omega$, is independent of the mass of the block.

Consider next the problem of quasi-static anti-plane sliding of identical isotropic, elastic continua (Rice and Ruina, 1983). Define a coordinate system such that the interface is located at $x_2 = 0$, the $x_1$ coordinate lies along the interface and slip occurs in the $x_3$ direction. Let $u_3(x_1, x_2, t)$ denote the displacement field. The governing equation for quasi-static anti-plane deformation is

$$\frac{\partial^2 u_3}{\partial x_1^2} + \frac{\partial^2 u_3}{\partial x_2^2} = 0 \qquad (16)$$

As before, we consider a state of steady sliding at velocity $V_o$, shear stress $\tau_o$ and normal stress $\sigma_o$. A displacement field which represents the steady sliding and a perturbation from it is

$$u_3(x_1, x_2, t) = \frac{1}{2} V_o t \, \text{sgn}(x_2) + \frac{1}{2} D(k,t) e^{ikx_1} e^{-|k||x_2|} \text{sgn}(x_2) \qquad (17)$$

where the first term denotes the steady sliding and the second term, the perturbation, satisfies the governing Laplace equation. The slip velocity perturbation from steady sliding is then

$$V(x_1, t) - V_o = \frac{\partial D(k,t)}{\partial t} e^{ikx_1} \qquad (18)$$

and the relation for the shear stress perturbation $\tau - \tau_o = \mu \partial u_3 / \partial x_2$, where $\mu$ is the shear modulus, gives

$$\dot{\tau} = -\frac{|k|\mu}{2}(V - V_o). \qquad (19)$$

From eqns (1) and (19) we see that $|k|\mu/2$ acts as the stiffness of a continuum in quasi-static anti-plane sliding. It follows from eqn (12) that the critical wavenumber for stability in anti-plane sliding of identical isotropic solids is

$$|k|_{cr} = \frac{2(b-a)\sigma_o}{\mu L}. \qquad (20)$$

Higher wavenumbers (i.e. shorter wavelengths) are stable and smaller wavenumbers (i.e. longer wavelengths) are unstable. The space-time dependence at the critical wavenumber has the form

$$V(x_1,t) - V_o : e^{\pm i|k|_{cr}(x_1 \pm ct)} \qquad (21)$$

where $c$, the phase velocity of the neutrally propagating mode is

$$c = \frac{\mu V_o}{2\sqrt{a(b-a)\sigma_o}}. \qquad (22)$$

It is noteworthy that the time dependence of the neutral mode is

$$V(x_1,t) - V_o : e^{\pm i\omega t} \qquad (23)$$

where

$$\omega = |k|_{cr} c = \sqrt{\frac{b-a}{a}} \frac{V_o}{L} \qquad (24)$$

is independent of elastic parameters. The corresponding result (Rice et. al., 2001) for dynamic sliding of the identical, isotropic half-spaces under anti-plane strain is

$$|k|_{cr} = \frac{2(b-a)\sigma_o}{\mu L}\sqrt{1+q^2}$$
$$c = qc_S / \sqrt{1+q^2} \qquad (25)$$

where $c_S$ is the shear wave speed of the solid and $q$ is a non-dimensional sliding velocity given by

$$q = \frac{\mu V_o}{2\sqrt{(b-a)a\sigma_o c_s}}.  \qquad (26)$$

It must be noted that the relation eqn. (24) still holds in the dynamic case, showing that the time dependence of the neutral mode is independent of inertial parameters also. Sliding is always stable with a velocity-strengthening friction law, $(b-a) < 0$.

The sliding of dissimilar materials has been a topic of much recent interest. When the solids on either side of a sliding interface have different elastic properties, non-uniform slip on the interface causes changes to the normal stress $\sigma_o$ acting on the interface. This has been shown to be inherently destabilizing both under quasi-static and dynamic conditions. For isotropic solids, the normal stress alteration occurs only when the spatial variation of the slip perturbation is in the direction of sliding (commonly called in-plane sliding) and not when it is transverse to the slip direction (the anti-plane sliding case discussed above). Under quasi-static conditions, Rice et. al. (2001) show that a slip velocity perturbation of the form given in eqn. (18) (but now with $x_1$ being the sliding direction instead of $x_3$) results in shear and normal stress changes

$$\tau - \tau_o = -\frac{|k|M}{2} D(k,t) e^{ikx_1}$$
$$\sigma - \sigma_o = \frac{ikM\beta}{2} D(k,t) e^{ikx_1},$$

where $M > 0$ is an elastic constant and $\beta$ is a Dundurs parameter. Rice et. al. (2001) show that due to the normal stress coupling, slip can be unstable even with a velocity-strengthening friction law, $(b-a) < 0$. When elastodynamic effects are included, Renardy (1992), Adams (1995), Martins and Simões (1995), Simões and Martins (1998) and Ranjith and Rice (2001) show that the steady sliding of dissimilar materials is often mathematically ill-posed due to a short-wavelength instability when a Coulomb friction law, $\tau = f\sigma$, where $f$ is a constant friction coefficient, acts on the interface. The time dependence of the slip response for an $\exp(ikx_1)$ perturbation is found to be of the form $\exp(\alpha |k| t)$ with $\alpha > 0$. Ranjith and Rice (2001) and Rice et. al. (2001) discuss corrections to the Coulomb law which can regularize the steady sliding problem. They show that a friction law with memory of normal stress history, as suggested by the experiments of Prakash and Clifton (1992) and Prakash (1998), of the form

$$\dot{\tau} = -\frac{V_o}{L}(\tau - f\sigma)$$

or a rate- and state-dependent law of the form eqn. (6) with $a > 0$ correct the short-wavelength instability.

The effect of elastic anisotropy on frictional instabilities has recently been studied by Ranjith and Gao (2007) for anti-plane as well as for in-plane sliding assuming quasi-static conditions. Results were specialized for the sliding of an orthotropic solid on an isotropic solid. In anti-plane shear, the orthotropic solid has two independent elastic constants, $C_{55}$ and $C_{44}$, and follows the constitutive laws

$$\tau_{31} = C_{55} \frac{\partial u_3}{\partial x_1},$$
$$\tau_{32} = C_{44} \frac{\partial u_3}{\partial x_2}, \quad (27)$$

where $\tau_{31}$ and $\tau_{32}$ are the anti-plane stresses. Ranjith and Gao (2007) showed that the critical wavenumber for stability in that case is

$$|k|_{cr} = \frac{\sigma_o(b-a)}{L} \frac{(1 + \mu/\sqrt{C_{55}C_{44}})}{\mu} \quad (28)$$

where $\mu$ is the shear modulus of the isotropic half-space. The phase velocity, $c$, of the neutral mode can be determined from eqn. (22), which still holds. An expression for the critical wavenumber under in-plane sliding was also obtained. It was found that for a range of orientations of the orthotropic solid, sliding is stable to perturbations of all wavelengths even with a velocity weakening friction law. Conversely, for the same range of orientations sliding becomes unstable even with a velocity strengthening friction law.

In the present paper, the work of Ranjith and Gao (2007) is extended to study the stability of dynamic, anti-plane sliding of dissimilar, anisotropic elastic solids having a plane of symmetry normal to the slip direction. Each anisotropic solid is characterized by three independent elastic constants (in anti-plane shear) and the density, but it is shown that the material constants enter the problem only in the form of an effective shear modulus and a characteristic wave speed, thus simplifying the analysis. Plots of the critical wavelength for stability, $|k|_{cr}$, and the phase velocity of the neutral mode, $c$, as a function of the non-dimensional sliding velocity, $q$, are obtained for various sets of bi-material parameters.

**2. Formulation**

Consider the origin of a Cartesian coordinate system placed at the planar interface between two unbounded dissimilar anisotropic solids. Let the $x_1$ coordinate lie along the interface. The anisotropic elastic solid occupying the region $x_2 \geq 0$ is labeled Material #1 while the solid occupying the region $x_2 \leq 0$ is labeled Material #2. It is assumed that the displacements $u_i$, $(i = 1, 2, 3)$ and the stresses $\tau_{ij}$, $(i, j = 1, 2, 3)$ are independent of the $x_3$ coordinate.

The theory of elasticity for a solid of arbitrary anisotropy, with the elastic fields being independent of one spatial coordinate, differs in important ways from isotropic elasticity (see, for example, Eshelby et. al., 1953 and Teutonico, 1962). For isotropic solids, it is well known that, when the elastic fields are invariant with respect to the $x_3$ direction, the stress components in the $x_1 - x_2$ plane depend only on the derivatives of $u_1$ and $u_2$ and the stress components in the $x_3$ direction depend only on the derivatives of $u_3$. For a straight edge dislocation along the $x_3$ axis in an isotropic solid, $u_3 = 0$, and for a straight screw dislocation along the $x_3$ axis, $u_1 = u_2 = 0$. This, however, is not true in general for anisotropic solids. Both edge and screw dislocations involve all three displacement components and all stress components. Therefore, changes to the normal stress on the sliding plane could be expected even in anti-plane sliding, which involves dislocations of the pure screw type. As mentioned in the Introduction, the coupling of slip to normal stress changes often gives rise to unexpected instabilities. An analysis of anti-plane sliding of dissimilar solids of arbitrary anisotropy therefore promises to be of considerable theoretical and practical interest. Eshelby et. al. (1953) also show that only when the $x_1 - x_2$ plane is a plane of symmetry, does a pure edge dislocation have $u_3 = 0$ and a pure screw dislocation have $u_1 = u_2 = 0$. The in-plane and anti-plane stress components also decouple in this case and this greatly simplifies the study of such solids.

In this paper, we consider the anti-plane sliding (in the $x_3$ direction) of two anisotropic solids with the $x_1 - x_2$ plane being a plane of symmetry. As noted above, the assumption of symmetry implies that only the $u_3$ displacement component is present and that normal stress on the interface is not affected by slip. In the steady state, the slip velocity is a constant, $V_o$, and a constant remote normal stress $\sigma_o$ and shear stress $\tau_o$ are applied so that $\tau_o = f\sigma_o$, where $f$ is the friction coefficient at the slip velocity $V_o$.

We first develop the elastodynamic relation between slip and traction perturbations at the interface. The derivation follows the method of Geubelle and Rice (1995) for isotropic solids. Consider a displacement field for anti-plane motions given by

$$u_1 = u_2 = 0,$$
$$u_3 = \frac{1}{2}V_o t \, \text{sgn}(x_2) + u(x_1, x_2, t). \tag{29}$$

where $u(x_1, x_2, t)$ represents the perturbation from steady sliding in the $x_3$ direction. The slip velocity perturbation from steady sliding is then given by

$$V - V_o = \frac{\partial u(x_1, x_2 = 0^+, t)}{\partial t} - \frac{\partial u(x_1, x_2 = 0^-, t)}{\partial t} \tag{30}$$

The stress perturbation in Material #1 corresponding to the displacement perturbation $u(x_1, x_2, t)$ is

$$\sigma_{31} = C_{55}\frac{\partial u}{\partial x_1} + C_{45}\frac{\partial u}{\partial x_2},$$
$$\sigma_{32} = C_{45}\frac{\partial u}{\partial x_1} + C_{44}\frac{\partial u}{\partial x_2}. \tag{31}$$

where $C_{44}, C_{45}, C_{55}$ are the elastic stiffnesses of Material #1, so that the traction perturbation on the plane $x_2 = 0$ is given by

$$\tau - \tau_o = \sigma_{32}(x_1, x_2 = 0^+, t) \tag{32}$$

The only non-trivial linear momentum balance equation for the perturbation stress and displacement fields is

$$\frac{\partial \sigma_{31}}{\partial x_1} + \frac{\partial \sigma_{32}}{\partial x_2} = \rho \frac{\partial^2 u}{\partial t^2} \tag{33}$$

where $\rho$ is the density of Material #1. Substituting for the stress field, eqn. (31) into eqn. (33), one gets the equation of motion,

$$C_{55} \frac{\partial^2 u}{\partial x_1^2} + 2C_{45} \frac{\partial^2 u}{\partial x_1 \partial x_2} + C_{44} \frac{\partial^2 u}{\partial x_2^2} = \rho \frac{\partial^2 u}{\partial t^2} \tag{34}$$

Let the Laplace transform be defined by

$$\hat{f}(p) = \int_0^\infty e^{-pt} f(t) dt . \tag{35}$$

Taking the Laplace transform of eqn. (34), we get

$$C_{55} \frac{\partial^2 \hat{u}}{\partial x_1^2} + 2C_{45} \frac{\partial^2 \hat{u}}{\partial x_1 \partial x_2} + C_{44} \frac{\partial^2 \hat{u}}{\partial x_2^2} = \rho p^2 \hat{u} \tag{36}$$

Consider now a single Fourier mode in the displacement of the form

$$u(x_1, x_2, t) = U(k, x_2, t) e^{ikx_1} . \tag{37}$$

Taking Laplace transform of eqn. (37) and substituting into eqn. (36), we get

$$-C_{55} k^2 \hat{U} + 2ikC_{45} \frac{\partial \hat{U}}{\partial x_2} + C_{44} \frac{\partial^2 \hat{U}}{\partial x_2^2} = \rho p^2 \hat{U} . \tag{38}$$

Noting that the above equation is an ordinary differential equation for $\hat{U}$ with constant coefficients, we look for a solution in the form

$$\hat{U}(x_2, k, p) = \hat{U}^+(k, p) e^{\alpha x_2} \tag{39}$$

Substituting into eqn. (38) one gets for $\alpha$,

$$\alpha^2 C_{44} + 2ikC_{45}\alpha - (\rho p^2 + C_{55}k^2) = 0. \tag{40}$$

This quadratic equation has two solutions; picking the root that gives a bounded displacement in the half-space $x_2 \geq 0$, we have

$$\alpha = \frac{-ikC_{45} - \sqrt{k^2(C_{44}C_{55} - C_{45}^2) + C_{44}\rho p^2}}{C_{44}} \tag{41}$$

The displacement perturbation at $x_2 = 0^+$ is then given by

$$u(x_1, x_2 = 0^+, t) = U^+(k,t)e^{ikx_1}. \tag{42}$$

The perturbation of the traction component of stress at $x_2 = 0^+$ is

$$\tau - \tau_o = C_{45}\frac{\partial u}{\partial x_1}(x_1, x_2 = 0^+, t) + C_{44}\frac{\partial u}{\partial x_2}(x_1, x_2 = 0^+, t). \tag{43}$$

Denoting the traction component of the stress perturbation on the plane $x_2 = 0$ by

$$\tau(x_1, t) - \tau_o \equiv T(k,t)e^{ikx_1}, \tag{44}$$

eqns. (42), (43) and (44) imply

$$\hat{T}(k,p) = -\sqrt{k^2(C_{44}C_{55} - C_{45}^2) + C_{44}\rho p^2}\,\hat{U}^+(k,p). \tag{45}$$

Writing

$$\mu = \sqrt{C_{44}C_{55} - C_{45}^2},\ c_1^2 = \frac{\mu^2}{C_{44}\rho}, \tag{46}$$

we can rewrite eqn. (45) as

$$\hat{T}(k,p) = -\mu|k|\sqrt{1 + p^2/k^2 c_1^2}\,\hat{U}^+(k,p). \tag{47}$$

It is of interest to note that the corresponding equation between traction and displacement perturbations for an isotropic half-space has a similar form as eqn. (47), with $\mu$ identified as the shear modulus of the half-space and $c_1$ as the shear wave speed.

A similar analysis of Material #2 occupying the half-space $x_2 \leq 0$ follows. The constitutive relations for the lower half-space are

$$\sigma_{31} = C'_{55} \frac{\partial u}{\partial x_1} + C'_{45} \frac{\partial u}{\partial x_2},$$
$$\sigma_{32} = C'_{45} \frac{\partial u}{\partial x_1} + C'_{44} \frac{\partial u}{\partial x_2}. \tag{48}$$

where $C'_{44}, C'_{45}, C'_{55}$ are the elastic stiffnesses of Material #2. The linear momentum balance equation for Material #2 is

$$\frac{\partial \sigma_{31}}{\partial x_1} + \frac{\partial \sigma_{32}}{\partial x_2} = \rho' \frac{\partial^2 u}{\partial t^2} \tag{49}$$

where $\rho'$ is the density of Material #2. We look for a displacement field perturbation in the lower half-space of the form

$$u(x_1, x_2, t) = U^-(k,t) e^{ikx_1} e^{\alpha' x_2}, \tag{50}$$

so that the displacement perturbation at $x_2 = 0^-$ is given by

$$u(x_1, x_2 = 0^-, t) = U^-(k,t) e^{ikx_1}. \tag{51}$$

Taking Laplace transform of eqn. (50), substituting into the equation of motion for the lower half-space and requiring that the displacements be bounded, one gets

$$\alpha' = \frac{-ikC'_{45} + \sqrt{k^2(C'_{44}C'_{55} - C'_{45}{}^2) + C'_{44}\rho' p^2}}{C'_{44}} \tag{52}$$

The perturbation of the traction component of stress is given by

$$\tau(x_1,t) - \tau_o = T(k,t)e^{ikx_1} = C'_{45}\frac{\partial u}{\partial x_1}(x_1, x_2 = 0^-, t) + C'_{44}\frac{\partial u}{\partial x_2}(x_1, x_2 = 0^-, t) \qquad (53)$$

which gives

$$\hat{T}(k,p) = +\mu'|k|\sqrt{1 + p^2/k^2 c_1'^2}\,\hat{U}^-(k,p), \qquad (54)$$

where

$$\mu' = \sqrt{C'_{44}C'_{55} - C'_{45}{}^2},\; c_1'^2 = \frac{\mu'^2}{C'_{44}\rho'}. \qquad (55)$$

In the following, we assume without loss of generality, that $c_1 < c_1'$.

We now seek the elastodynamic relation between the slip and the traction perturbations at the bi-material interface. From eqn. (42) and eqn. (51), we may write

$$u(x_1, x_2 = 0^+, t) - u(x_1, x_2 = 0^-, t) = (U^+(k,t) - U^-(k,t))e^{ikx_1} \equiv D(k,t)e^{ikx_1} \qquad (56)$$

so that the slip velocity perturbation defined by eqn. (30) is

$$V - V_o = \frac{\partial D(k,t)}{\partial t} e^{ikx_1} \qquad (57)$$

Using traction continuity at the interface, we may combine eqn. (47) and eqn. (54) in the form

$$\hat{T}(k,p) = -\frac{\mu|k|}{2} F(k,p)\hat{D}(k,p) \qquad (58)$$

where

$$F(k,p) = \frac{2\mu'\sqrt{1 + p^2/k^2 c_1^2}\sqrt{1 + p^2/k^2 c_1'^2}}{\mu\sqrt{1 + p^2/k^2 c_1^2} + \mu'\sqrt{1 + p^2/k^2 c_1'^2}}. \qquad (59)$$

When the two sliding solids are isotropic and have the same elastic properties and densities, eqn. (58) reduces to

$$\hat{T}(k,p) = -\frac{\mu|k|}{2}\sqrt{1 + p^2/k^2 c_1^2}\,\hat{D}(k,p), \qquad (60)$$

where $\mu$ is the shear modulus and $c_1 = \sqrt{\mu/\rho}$ is the shear wave speed of the solids, in agreement with the results of Geubelle and Rice (1995). For the case of quasi-static sliding (i.e. with inertial effects being ignored) of dissimilar anisotropic solids, eqn. (58) gives

$$\hat{T}(k,p) = -\frac{\mu\mu'}{\mu + \mu'}|k|\hat{D}(k,p). \qquad (61)$$

As observed earlier, it is of interest that the anisotropic elastic constants enter the problem only in the form of the combinations $\mu$ and $\mu'$.

We now derive the governing equation for slip stability when a rate and state dependent friction law is operative on the interface. Taking the Laplace transform of the linearized friction law, eqn. (10), we obtain the relation between the traction perturbation $\tau - \tau_o = T(k,t)e^{ikx_1}$ and slip velocity perturbation $V - V_o = \frac{\partial D(k,t)}{\partial t}e^{ikx_1}$ as

$$\left(p + \frac{V_o}{L}\right)\hat{T} = \frac{\sigma_o}{V_o}\left[ap - (b-a)\frac{V_o}{L}\right]p\hat{D}. \qquad (62)$$

Requiring that the traction and slip velocity perturbations also satisfy the elastodynamic relation, eqn. (58), one obtains the equation governing the stability of sliding as

$$\frac{\mu}{2}\left(p + \frac{V_o}{L}\right)|k|F(k,p) + \frac{\sigma_o p}{V_o}\left[ap - (b-a)\frac{V_o}{L}\right] = 0. \qquad (63)$$

This equation has the same form as that obtained in Rice et. al. (2001) for dynamic sliding between identical isotropic solids. The function $F(k,p)$ is more complicated expectedly for the dissimilar material case. The analysis of eqn. (63) presented below largely follows that laid out in Rice et. al. (2001). It is clear that for a given value of $|k|$, the roots $p$ of eqn. (63) occur in complex conjugate pairs. If the roots have a positive real part for a given $|k|$, then a perturbation of wavelength $\lambda = 2\pi/|k|$ is unstable. It may be shown, following Rice and Ruina (1983), that for large $|k|$, the real part of the roots is negative when $a > 0$. As $|k|$ is reduced from large values, stability is lost for velocity weakening surfaces, $b - a > 0$, by way of a Hopf bifurcation. At the Hopf bifurcation, a pair of complex conjugate roots cross the imaginary $p$-axis from the $\text{Re}(p) < 0$ domain to the $\text{Re}(p) > 0$ domain as $|k|$ is decreased below a critical value, $|k|_{cr}$. We now determine $|k|_{cr}$ and the value of $p$ at the Hopf bifurcation. To do so, we write the value of $p$ at the Hopf bifurcation as $p = \pm i |k| c$ and determine all values of $|k|$ and $c$ (and hence of $p$) at which the bifurcation occurs. The bifurcation that gives the largest value of $|k|$ corresponds to $|k|_{cr}$, since it the value at which stability is lost as $|k|$ is reduced from large values. In the following, we focus on the root of the form $p = +i |k| c$ with $c > 0$ and looks for imaginary $p$-axis crossings successively in the region $0 < c < c_1$, $c_1 < c < c_1'$ and $c_1' < c$.

First, we consider the case $0 < c < c_1$. Substituting $p = +i|k|c$ in eqn. (23) and separating the real and imaginary parts of the resulting equation, we get the pair of equations

$$\frac{V_o}{L}\frac{\mu}{2}F(c) = \frac{a\sigma_o}{V_o}|k|c^2,$$
$$\frac{\mu|k|}{2}F(c) = \frac{(b-a)\sigma_o}{L}. \tag{64}$$

where

$$F(c) = \frac{2\mu'\sqrt{1-c^2/c_1^2}\sqrt{1-c^2/c_1'^2}}{\mu\sqrt{1-c^2/c_1^2} + \mu'\sqrt{1-c^2/c_1'^2}} \tag{65}$$

Dividing the two equations in eqn. (64) by one another, we get

$$|k|c = \sqrt{\frac{b-a}{a}}\frac{V_o}{L}. \tag{66}$$

Substituting eqn. (66) in the first of eqn. (64), one obtains,

$$\frac{c/c_1}{F(c)} = \frac{\mu V_o}{2\sqrt{a(b-a)}\sigma_o c_1} \equiv q. \tag{67}$$

This equation can be solved numerically for $c$ for a given value of the non-dimensional sliding velocity $q$. The value of $|k|$ corresponding to the value of $q$ can then be found from eqn. (66).

When $c_1 < c < c_1'$, $F(c)$ is a complex number. Writing $F(c) = F_1(c) + iF_2(c)$, where $F_1(c)$ and $F_2(c)$ are real, following the analysis in Rice et. al. (2001), we can obtain relations analogous to eqn. (66) and eqn. (67) when $c_1 < c < c_1'$ as

$$|k|c = \left[\sqrt{\left[\frac{bF_2(c)}{2aF_1(c)}\right]^2 + \frac{b-a}{a}} - \frac{bF_2(c)}{2aF_1(c)}\right]\frac{V_o}{L} \tag{68}$$

and

$$\frac{\sqrt{\frac{b}{a}-1}\frac{c}{c_1}}{\sqrt{\left[\frac{F_2(c)b}{2a}\right]^2 + \left(\frac{b}{a}-1\right)F_1^2(c)} - \frac{F_2(c)b}{2a} + F_2(c)} = \frac{\mu V_o}{2\sqrt{a(b-a)}\sigma_o c_1} \equiv q. \tag{69}$$

Finally, it can be shown that no Hopf bifurcations of eqn. (63) occur when $c > c_1'$.

## 3. Discussion

Figs. 1 to 8 shows the dependence of the wavenumber magnitude, $|k|$, at the Hopf bifurcations obtained from eqn. (66) and eqn. (68) and the corresponding phase velocity, $c$, of the neutrally propagating mode obtained from eqn. (67) and eqn. (69) as a function of the non-dimensional sliding velocity, $q$, for surfaces that are velocity-weakening in the steady state, $b - a > 0$, and for bi-material combinations of interest in earthquake mechanics, technology and biology.

Figs. 1 and 2 are for a bi-material system with $c_1' = 1.2 c_1$ and $\mu'/\mu = 1$. Such a low contrast in material properties often occurs across active faults in the earth. The upper curve in Fig. 1 corresponds to the mode with $c < c_1$ (for all values of $b/a$), while the lower curve is for $c_1 < c < c_1'$ with $b/a = 1.2$. Clearly, the upper curve yields the critical wavenumber, $|k|_{cr}$, since it has a higher value of $|k|$ for a given value of $q$. The phase velocity, $c$, corresponding to $|k|_{cr}$ thus satisfies $c < c_1$. It is of interest to note that for a range of values of $q$ there are two additional neutrally propagating modes with phase velocity in the range $c_1 < c < c_1'$. The results indicate the possibility of anti-plane rupture propagation at a bi-material interface with a speed greater than the shear wave speed of a constituent material and could have important implications for studies of earthquake processes and dynamic fractures in technological and biological systems. Figs. 3 to 8 show similar plots of $|k|$ and $c$ at the Hopf bifurcation for bi-material systems with a large contrast in the characteristic wave speeds and effective shear moduli, as is typical of interfaces in technological and biological systems. The value $c_1' = 5 c_1$ is chosen for all plots in Figs. 3 to 8 with $\mu'/\mu$ taking the values 1 (Figs. 3 and 4), 10 (Figs. 5 and 6) and 0.1 (Figs. 7 and 8). Figs. 3 to 8 show qualitative features similar to Figs. 1 and 2.

As noted in the Formulation section, the results strictly hold only when the plane normal to the slip direction is a plane of symmetry. In the general case, anti-plane sliding will also cause in-plane displacements and normal stress changes on the slip plane. An analysis of the more general case should be of considerable interest.

**4. Conclusions**:

The stability to perturbations from steady dynamic anti-plane sliding with velocity $V_o$ at an interface between dissimilar anisotropic materials has been studied. The plane normal to the direction of sliding is assumed to be a plane of elastic symmetry. Friction at the interface is assumed to follow a rate and state dependent constitutive law with velocity-weakening behavior in the steady state. Each anisotropic material is characterized by three elastic constants (in anti-plane shear) and the density but it is found that the material properties of the two solids enter the problem in combinations of effective shear moduli, $\mu$ and $\mu'$, and characteristic wave speeds, $c_1$ and $c'_1$. The perturbations at the interface of the materials have a spatial dependence of the form $\exp(ikx_1)$ and the resulting neutrally stable slip rate modes of the form $V - V_o \sim \exp[i|k|(x_1 \pm ct)]$ are determined. The values of the wavenumber magnitude, $|k|$, and phase velocity, $c$, at neutral stability are determined as a function of $V_o$. The critical wavenumber magnitude, $|k|_{cr}$, above which there is stability to perturbation and the corresponding phase velocity, $c$, are hence determined. It is found that the phase velocity corresponding to $|k|_{cr}$ satisfies $c < \min(c_1, c_1')$. It is also observed that two additional modes of neutrally stable slip exist for sufficiently large values of $V_o$, with phase velocity in between the characteristic wave speeds of the two materials, $c_1 < c < c_1'$.

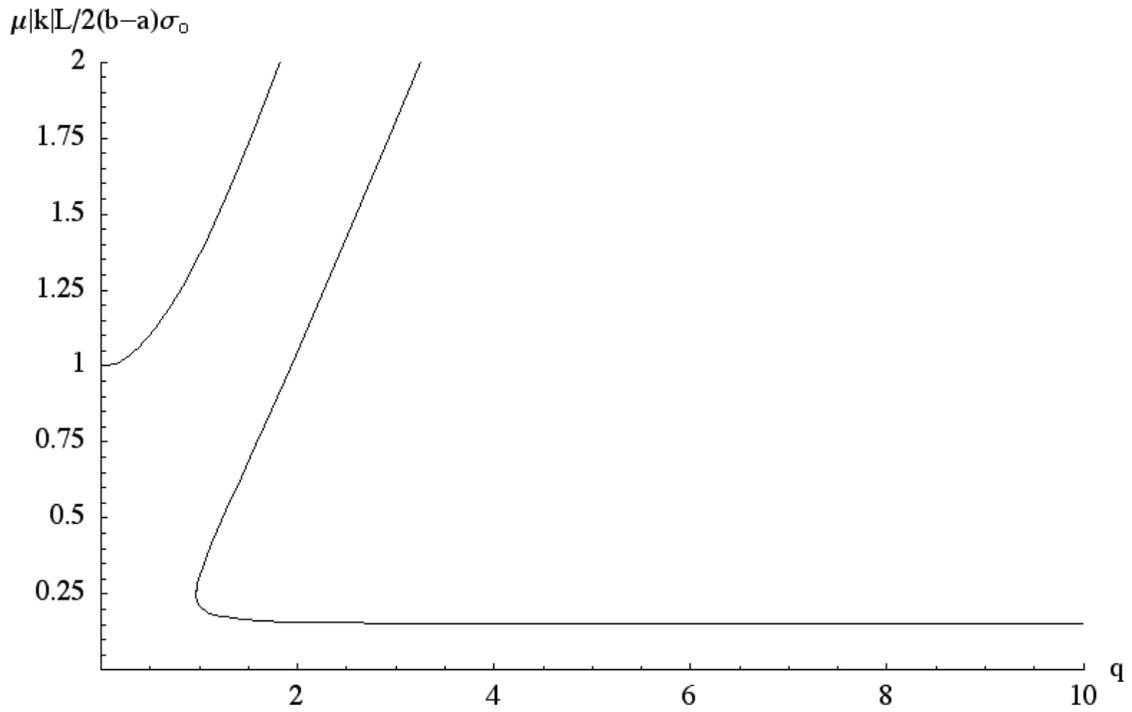

Figure 1: Normalized wavenumber magnitude, $|k|$, at Hopf bifurcation as a function of non-dimensional unperturbed sliding velocity, $q$, for a bi-material system with $c_1' = 1.2 c_1$ and $\mu' = \mu$. The upper curve holds for all values of $b/a$ while the lower curve is for $b/a = 1.2$.

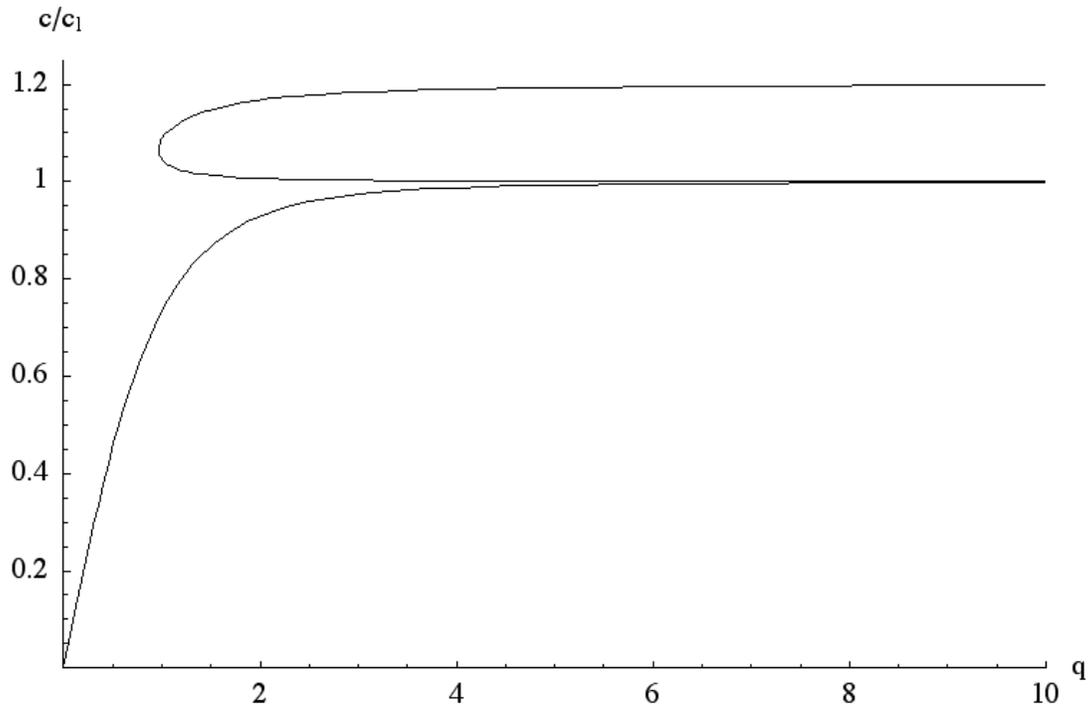

Figure 2: Normalized phase velocity, $c$, at Hopf bifurcation as a function of non-dimensional unperturbed sliding velocity, $q$, for a bi-material system with $c_1' = 1.2 c_1$ and $\mu' = \mu$. The lower curve holds for all values of $b/a$ while the upper curve is for $b/a = 1.2$.

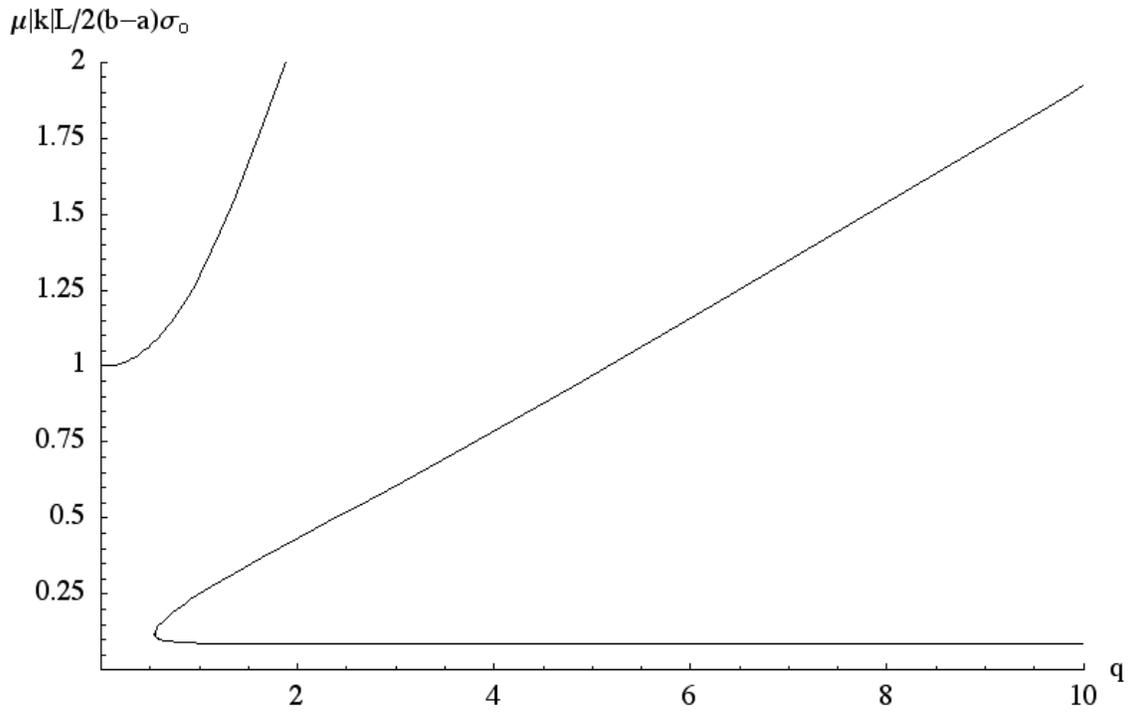

Figure 3: Normalized wavenumber magnitude, $|k|$, at Hopf bifurcation as a function of non-dimensional unperturbed sliding velocity, $q$, for a bi-material system with $c_1' = 5c_1$ and $\mu' = \mu$. The upper curve holds for all values of $b/a$ while the lower curve is for $b/a = 1.2$.

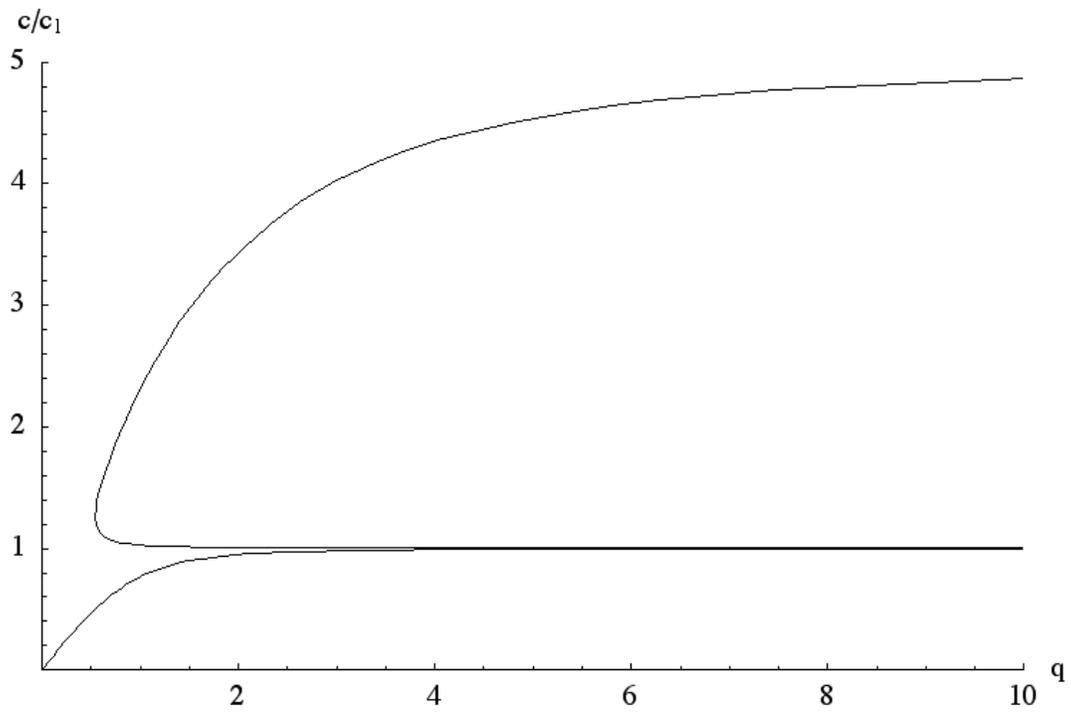

Figure 4: Normalized phase velocity, $c$, at Hopf bifurcation as a function of non-dimensional unperturbed sliding velocity, $q$, for a bi-material system with $c'_1 = 5c_1$ and $\mu' = \mu$. The lower curve holds for all values of $b/a$ while the upper curve is for $b/a = 1.2$.

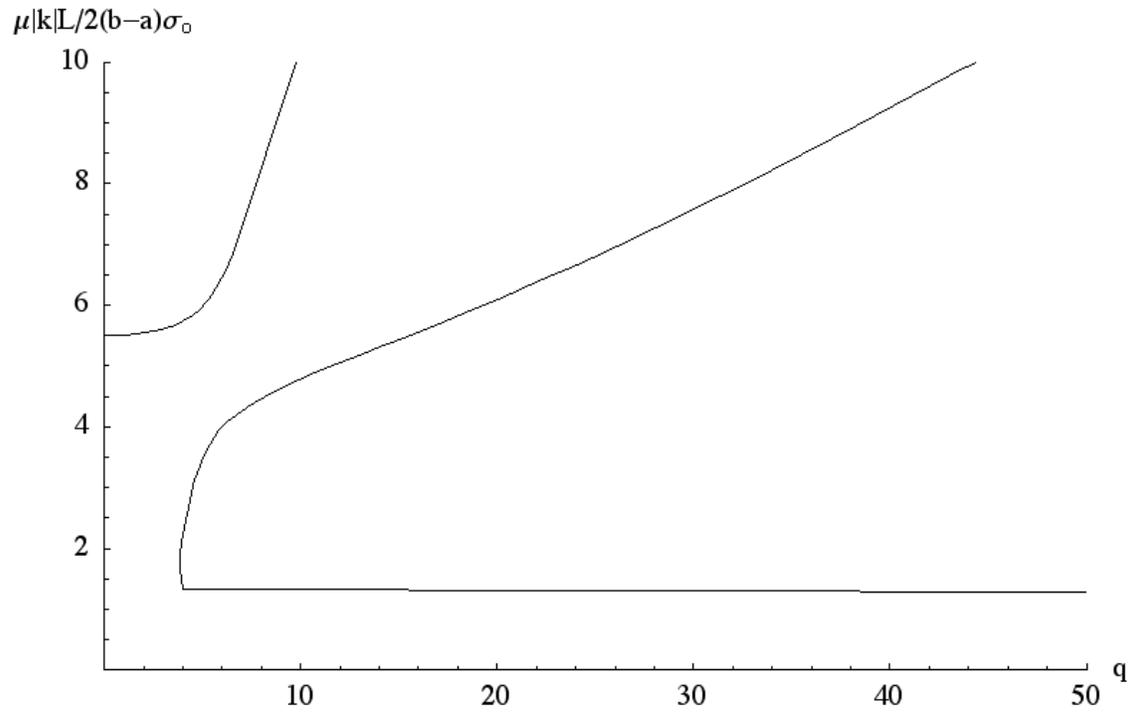

Figure 5: Normalized wavenumber magnitude, $|k|$, at Hopf bifurcation as a function of non-dimensional unperturbed sliding velocity, $q$, for a bi-material system with $c_1' = 5c_1$ and $\mu' = 10\mu$. The upper curve holds for all values of $b/a$ while the lower curve is for $b/a = 1.2$.

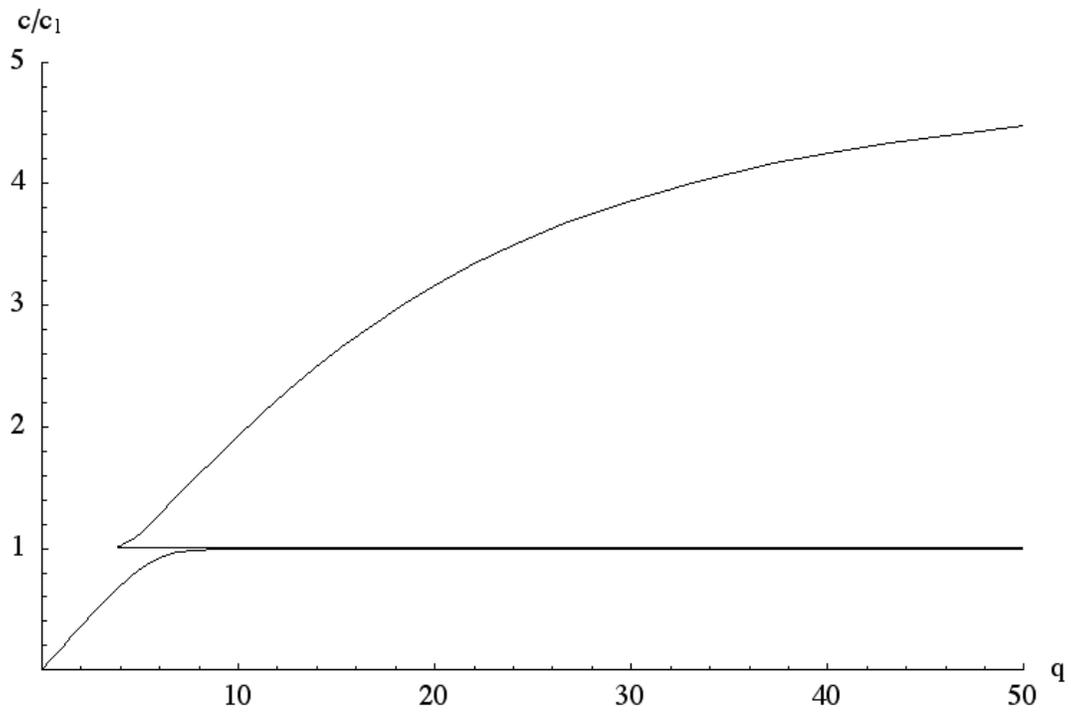

Figure 6: Normalized phase velocity, $c$, at Hopf bifurcation as a function of non-dimensional unperturbed sliding velocity, $q$, for a bi-material system with $c_1' = 5c_1$ and $\mu' = 10\mu$. The lower curve holds for all values of $b/a$ while the upper curve is for $b/a = 1.2$.

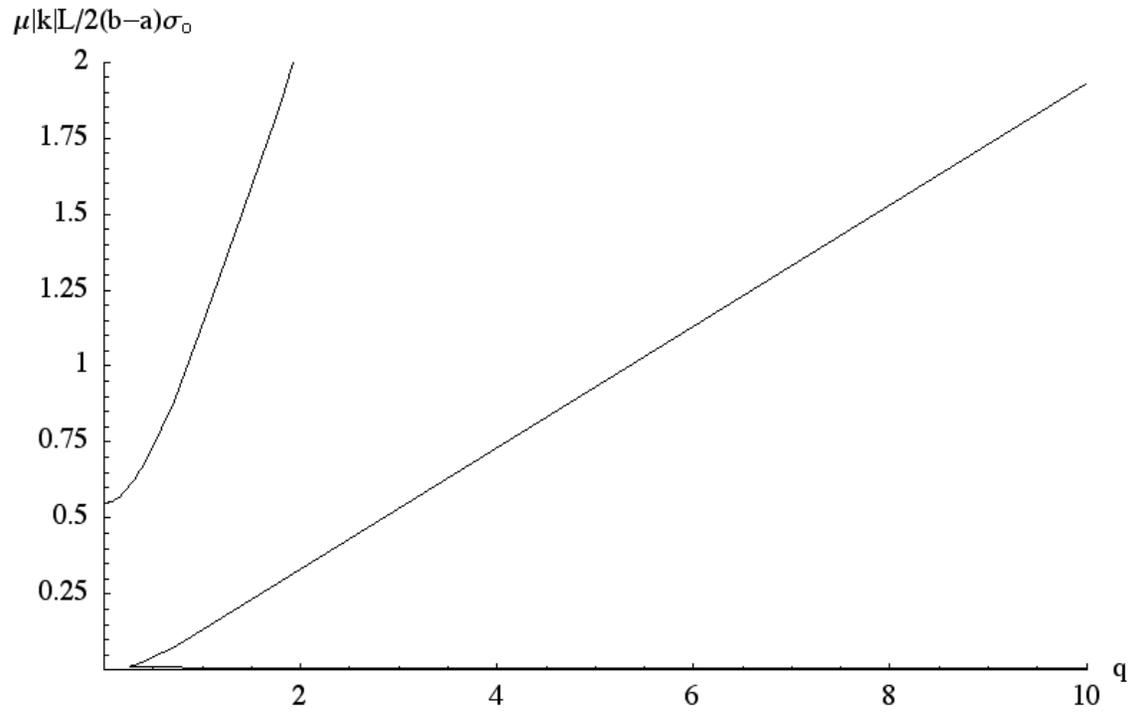

Figure 7: Normalized wavenumber magnitude, $|k|$, at Hopf bifurcation as a function of non-dimensional unperturbed sliding velocity, $q$, for a bi-material system with $c'_1 = 5c_1$ and $\mu' = 0.1\mu$. The upper curve holds for all values of $b/a$ while the lower curve is for $b/a = 1.2$.

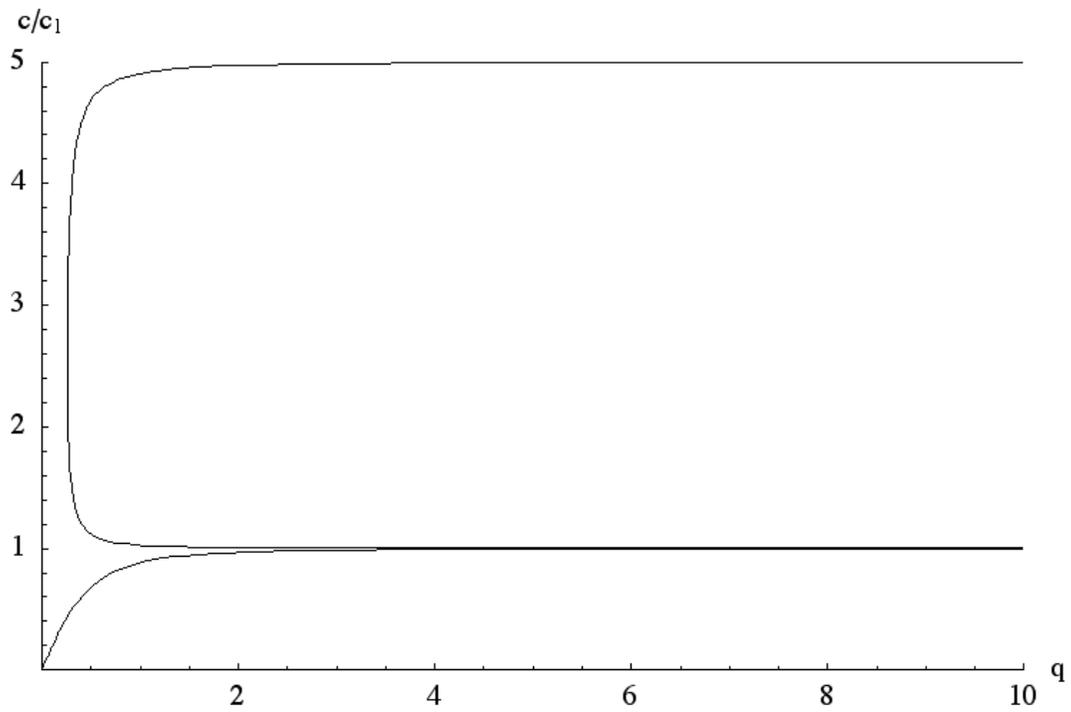

Figure 8: Normalized phase velocity, $c$, at Hopf bifurcation as a function of non-dimensional unperturbed sliding velocity, $q$, for a bi-material system with $c_1' = 5c_1$ and $\mu' = 0.1\mu$. The lower curve holds for all values of $b/a$ while the upper curve is for $b/a = 1.2$.

# References

Adams, G.G., 1995. Self-excited oscillations of two elastic half-spaces sliding with a constant coefficient of friction. J. Appl. Mech. 62, 867-872.

Dieterich, J.H., 1979. Modeling of rock friction – I. Experimental results and constitutive equations. J. Geophys. Res. 84, 2161-2168.

Eshelby, J. D., Read, W. T., Shockley, W., 1953. Anisotropic elasticity with applications to dislocation theory. Acta Metallurgica, 1, 251-259.

Geubelle, P.H., Rice, J.R., 1995. A spectral method for three-dimensional elastodynamic fracture problems. J. Mech. Phys. Solids 43, 1791-1824.

Martins, J.A.C., Simões. F.M.F., 1995. On some sources of instability/ill-posedness in elasticity problems with Coulomb friction. In: Raous, M., Jean, M., Moreau, J.J. (Eds.), Contact Mechanics. Plenum Press, New York, pp. 95-106.

Prakash, V., 1998. Frictional response of sliding interfaces subjected to time varying normal pressures. J. Tribol. Trans. ASME 120, 97-102.

Prakash, V., Clifton, R.J., 1992. Pressure-shear plate impact measurement of dynamic friction for high speed machining applications. Proceedings of the Seventh International Congress on Experimental Mechanics. Society of Experimental Mechanics, Bethel, Conn., pp. 556-564.

Ranjith, K., Rice, J.R., 2001. Slip dynamics at an interface between dissimilar materials. J. Mech. Phys. Solids 49, 341-361.

Ranjith, K., Gao, H., 2007. Stability of frictional slipping at an anisotropic/isotropic interface. Int. J. Solids Struct. 44, 4318-4328.

Renardy, M., 1992. Ill-posedness at the boundary for elastic solids sliding under Coulomb friction. J. Elasticity 27, 281-287.

Rice, J.R., 1983. Constitutive relations for fault slip and earthquake instabilities. Pure Appl. Geophys. 121, 443-475.

Rice, J.R., Ruina, A.L., 1983. Stability of steady frictional slipping. J. Appl. Mech. 50, 343-349.

Rice, J.R., Lapusta, N., Ranjith, K., 2001. Rate and state dependent friction and the stability of sliding between elastically deformable solids. J. Mech. Phys. Solids 49, 1865-1898.

Ruina, A.L., 1983. Slip instability and state variable friction laws. J. Geophys. Res. 88, 10359-10370.

Simões, F.M.F., Martins, J.A.C., 1998. Instability and ill-posedness in some friction problems. Int. J. Eng. Sci. 36, 1265-1293.

Teutonico, L. J., 1962. Uniformly moving dislocations of arbitrary orientation in anisotropic media. Phys. Rev. 127, 413-418.